\documentclass[12pt,twocolumn,twoside]{article}

\usepackage{titlesec}
\usepackage{fancyhdr}
\usepackage{authblk}
\usepackage{siunitx}
\usepackage{setspace,verbatim}
\usepackage{transparent}

\usepackage{xcolor,colortbl}
\usepackage{graphicx}
\usepackage{longtable}
\usepackage{booktabs}
\usepackage{multirow}
\usepackage[colorlinks, linkcolor=blue, anchorcolor=blue, citecolor=blue]{hyperref}
\usepackage{subcaption}
\usepackage{amsfonts}
\usepackage{amssymb}
\usepackage{amsmath}
\usepackage{lscape}
\usepackage{blindtext}
\usepackage{subcaption}
\usepackage{textcomp}
\usepackage{gensymb}
\usepackage{threeparttable}
\usepackage{booktabs, caption, makecell}
\usepackage{url}

\usepackage{breakurl}
\usepackage{hyperref}

\usepackage[top=1in,bottom=1in,left=1in,right=1in]{geometry}

\def\01{\ensuremath{0\mathord{-}1}}
\def\st{\mathop{\rm s.t.}}

\usepackage{algorithm}
\usepackage{algorithmicx}
\usepackage{algpseudocode}

\usepackage[customcolors]{hf-tikz}

\tikzset{style green/.style={
    set fill color=green!50!lime!60,
    set border color=white,
  },
  style cyan/.style={
    set fill color=cyan!90!blue!60,
    set border color=white,
  },
  style orange/.style={
    set fill color=orange!80!red!60,
    set border color=white,
  },
  hor/.style={
    above left offset={-0.15,0.31},
    below right offset={0.15,-0.125},
    #1
  },
  ver/.style={
    above left offset={-0.1,0.3},
    below right offset={0.15,-0.15},
    #1
  }
}

\titleformat*{\section}{\normalsize\bfseries\sffamily}
\titleformat*{\subsection}{\small\bfseries\sffamily}
\titleformat*{\subsubsection}{\small\bfseries\sffamily}

\usepackage[font=small,labelfont=bf]{caption}

\newcommand\shorttitle{Simulation optimization: A review of algorithms and applications}
\newcommand\authors{Amaran, Sahinidis, Sharda and Bury}

\fancypagestyle{firststyle}
{
   \fancyhf{}

   \fancyfoot[L]{\footnotesize\scshape \copyright 2017. This manuscript version is made available under the CC-BY-NC-ND 4.0 license \url{http://creativecommons.org/licenses/by-nc-nd/4.0/}.  The formal publication of this article is available at \url{https://link.springer.com/article/10.1007/s10479-015-2019-x}.}
}

\newcommand{\objective}[2]{\ensuremath{\underset{#1}\min & & #2\\[5pt]}}

\newcommand{\constraint}[2][]{& & #2 & #1}

\newenvironment{optimize}
{\begin{equation}
\begin{array}{rlll}}
{\end{array}
\end{equation}}

\title{\textbf{Simulation optimization: A review of algorithms and applications}}
\author[1,2]{Satyajith Amaran}
\author[1]{Nikolaos~V.~Sahinidis}
\author[3]{Bikram Sharda}
\author[4]{Scott J. Bury}
\affil[1]{Department of Chemical Engineering, Carnegie Mellon University, Pittsburgh, PA 15213, USA}
\affil[2]{Now at Core R\&D, The Dow Chemical Company, Freeport, TX 77541, USA}
\affil[3]{Core R\&D, The Dow Chemical Company, Freeport, TX 77541, USA}
\affil[4]{Core R\&D, The Dow Chemical Company, Midland, MI 48667, USA}
\date{}

\begin{document}

\begin{singlespacing}
\twocolumn[{%
\maketitle
\thispagestyle{firststyle}

\bibliographystyle{plainnat}


\footnotesize
\vspace{-1cm}
\begin{abstract}
Simulation Optimization (SO) refers to the optimization of an objective function subject to constraints, both of which can be evaluated through a stochastic simulation. To address specific features of a particular simulation---discrete or continuous decisions, expensive or cheap simulations, single or multiple outputs, homogeneous or heterogeneous noise---various algorithms have been proposed in the literature. As one can imagine, there exist several competing algorithms for each of these classes of problems. This document emphasizes the difficulties in simulation optimization as compared to mathematical programming, makes reference to state-of-the-art algorithms in the field, examines and contrasts the different approaches used, reviews some of the diverse applications that have been tackled by these methods, and speculates on future directions in the field.
\end{abstract}


\vspace*{0.8cm}

}]

\footnotesize

\section{Introduction}
\label{sec:intro}

Advances in modeling and availability of cheap computational power have enabled the science, engineering, and business research communities to make use of simulations to model phenomena and systems. It is only natural that there be a great interest in manipulating degrees of freedom in the simulations to optimize them.

The term Simulation Optimization (SO) is an umbrella term for techniques used to optimize stochastic simulations. Simulation optimization involves the search for those specific settings of the input parameters to a stochastic simulation such that a target objective, which is a function of the simulation output, is, without loss of generality, minimized.

As opposed to mathematical programming, SO does not assume that an algebraic description of the simulation is available---the simulation may be available as a black box that only allows the evaluation of the objective and constraints for a particular input. In fact, many SO algorithmic approaches solely depend on such input-output data from the simulation in their search for optimal input settings.

In addition, many large-scale and/or detailed simulations may be expensive to run, in terms of time, money, or resources. As a result, there is also a need to perform few simulations in this search for optimal parameters. Outputs from these stochastic simulations are not deterministic, and usually follow some output distribution, which may or may not vary across the parametric space. This uncertainty or variability in output also adds to the challenge of optimization, as it becomes harder to discern the quality of the parametric input in the presence of this output noise. In addition, when an algebraic description of the simulation is not accessible, derivative information is usually unavailable, and the estimation of derivatives from the use of finite differences may not be suitable due to noisy outputs and the expensive nature of simulations.

The nature of the stochastic simulations under study will determine the specific technique chosen to optimize them. The simulations, which are often discrete-event simulations, may be partially accessible to us in algebraic form, or may be purely available as an input-output model (as a black box); they may have single or multiple outputs; they may have deterministic or stochastic output(s); they may involve discrete or continuous parameters; and they may or may not involve explicit, or even implicit/hidden constraints.

A very general simulation optimization problem can be represented by (\ref{eq:gen_problem}).\\
\begin{optimize}
\label{eq:gen_problem}
\objective{}{\mathbb{E}_{\omega} [f(x,y,\omega)]}
\st \constraint{\mathbb{E}_{\omega}[ g(x,y,\omega)] \leq 0}{}\\[3pt]
\constraint{h(x,y) \leq 0}{} \\[3pt]
\constraint{x_l \leq x \leq x_u}{}\\
\constraint{x \in \mathbb{R}^n, y \in \mathbb{D}^m.}{}
\tag{P1}
\end{optimize}
The function $f$ can be evaluated through simulation for a particular instance of the continuous inputs $x$, discrete inputs $y$, and a realization of the random variables in the simulation, the vector $\omega$ (which may or may not be a function of the inputs, $x$ and $y$). Similarly, the constraints defined by the vector-valued function $g$ are also evaluated with each simulation run. In this formulation, expected values for these stochastic functions are used. There may be other constraints (represented by $h$) that do not involve random variables, as well as bound constraints on the decision variables.

The relaxation of any of these conditions would constitute a problem that would fall under the purview of SO. Most algorithms focus on problems that either have solely discrete choices, or solely continuous decisions to make. Each constraint may be thought of as representing additional outputs from the simulation that need to be taken into consideration. In addition, there may be bound constraints imposed on decision variables, that may either be available or obtained from domain-specific knowledge. Relatively few existing algorithms attempt to address both discrete and continuous choices simultaneously, although some broad classes of approaches naturally lend themselves to be applicable in either, and therefore both, settings. Further, the discrete variables may either be binary, integer-ordered, or categorical and lie in some discrete space $\mathbb{D}$. 

As can be seen, the formulation \ref{eq:gen_problem} is extremely general, and therefore a wide variety of applications fall under the scope of simulation optimization. Various applications of simulation optimization in diverse research fields are tabulated in Section~\ref{sec:applications}.

Another common assumption is that $f$ is a real-valued function and $g$ is a real vector-valued function, both of whose expected values  may or may not be smooth or continuous functions. The most common objective in SO is to optimize the expected value of some performance metric, but other objective functions may be appropriate depending on the application. For instance, an objective that minimizes risk could be a possible alternative, in which case one would incorporate some sort of variance measure as well into the objective. 

This paper is an updated version of Amaran et al. \cite{a+14_4or} and is meant to be a survey of available techniques as well as recent advances in simulation optimization. The remainder of the introduction section provides a literature survey of prior reviews, and elaborates on the relationship of simulation optimization to mathematical programming, derivative-free optimization, and machine learning. Section \ref{sec:applications} provides a glimpse into the wide variety of applications of simulation optimization that have appeared in the literature. Section \ref{sec:algos} focuses on various algorithms for discrete and continuous simulation optimization, provides basic pseudocode for major categories of algorithms, and provides comprehensive references for each type of algorithm. Section \ref{sec:software}  provides a listing of available software for simulation optimization and Section \ref{sec:comparisons} discusses means to compare their performance. Section \ref{sec:conclusions} summarizes the progress of the field, and outlines some current and future topics for research.

\subsection{Prior reviews of simulation optimization}
\label{sec:lit_rev}

Several review papers (cf. \cite{meketon87, js89, safizadeh90, azadivar92, fu94, cm97, andradottir98, azadivar99, shjs00, f+00, fu02, ts04, fga05, andradottir06a, hn09, ahcm11, pg13}), books and research monographs (cf. \cite{spall03,rk04,kleijnen08,cl10}), and theses (cf. \cite{angun04,driessen06,deng07,chang08,frazier09,kabirian09}) have traced the development of simulation optimization.

Meketon \cite{meketon87} provides a classification of algorithmic approaches for optimization over simulations based on how much information or structure about the underlying model is known. The paper surveys the progress of the field between 1975 and 1987, and focuses on continuous simulation optimization. Andrad\'{o}ttir \cite{andradottir98} provides a tutorial on gradient-based procedures for continuous problems. Carson and Maria \cite{cm97} and Azadivar \cite{azadivar99} also give brief outlines of and pointers to prevailing simulation optimization algorithms.

Fu et al. \cite{f+00} contains several position statements of eminent researchers and practitioners in the field of simulation, where the integration of simulation with optimization is discussed. The issues addressed include generality vs. specificity of an algorithm, the wider scope of problems that simulation optimization methodologies have the potential to address, and the need for integrating provably convergent algorithms proposed by the research community with metaheuristics often used by commercial simulation software packages.

Of the more recent surveys, Fu \cite{fu94} provides an excellent tutorial on simulation optimization, and focuses on continuous optimization problems more than discrete optimization problems. The paper focuses specifically on discrete-event simulations. Fu \cite{fu02} provides a comprehensive survey of the field and its scope---the paper outlines the different ways in which optimization and simulation interact, gives examples of real-world applications, introduces simulation software and the optimization routines that each of them use, provides a very basic tutorial on simulation output analysis and convergence theory for simulation optimization, elaborates on algorithms for both continuous and discrete problems, and provides pointers to many useful sources. Fu et al. \cite{fga05} provide a concise, updated version of all of this, and also talk about estimation of distribution algorithms.

Tekin and Sabuncuoglu \cite{ts04} provide a table that analyzes past review papers and the techniques they focus on. Apart from providing detailed updates on advances in approaches and algorithms, the paper also lists references that attempt to compare different SO techniques. Hong and Nelson \cite{hn09} classify simulation optimization problems into those with (1) a finite number of solutions; (2) continuous decision variables; and (3) discrete variables that are integer-ordered. The paper describes procedures for each of these classes. Perhaps the most recent survey, \cite{ahcm11}, classifies simulation optimization algorithms and provides a survey of methods as well as applications appearing in the literature between 1995 and 2010.

The present paper provides an overview of techniques, and briefly outlines well-established methods with pointers to more detailed surveys, while expounding on more recent methods in a concise manner. Though several reviews exist, we catalog the most recent developments---the emergence of derivative-free optimization and its relationship with simulation optimization, the appearance of simulation test-beds for comparing algorithms, the recent application of simulation optimization in diverse fields, the development of and interest in related techniques and theory by the machine learning community and the optimization community, as well as the sheer unprecedented nature of recent interest in optimizing over simulations. A reflection of a surge in recent interest is evidenced by the fact that more than half of the works we reference were published in the last decade. The intent is to not only trace the progress of the field, but to provide an update on state-of-the-art methods and implementations, point the familiar as well as the uninitiated reader to relevant sources in the literature, and to speculate on future directions in the field.

\subsection{A note on terminology and scope}

As simulation optimization involves the use of algorithms that arose from widely differing fields (Section \ref{sec:algos}), has relationships to many diverse disciplines (Section \ref{sec:relationship}), and has been applied to many different practical applications from biology to engineering to logistics (Section \ref{sec:applications}), it is not surprising that it is known by various names in different fields. It has also been referred to as simulation-based optimization, stochastic optimization, parametric optimization, black-box optimization, and Optimization via Simulation (OvS), where the continuous and discrete versions are accordingly known as Continuous Optimization via Simulation (COvS) and Discrete Optimization via Simulation (DOvS). Each algorithmic technique may also go by different names, and we attempt to reconcile these in Section \ref{sec:algos}.

Inputs to the simulation may be variously referred to as parameter settings, input settings, variables, controls, solutions, designs, experiments (or experimental designs), factors, or configurations. Outputs from the simulation are called measurements, responses, performance metrics, objective values, simulation replications, realizations, or results. The performance of a simulation may also be referred to as an experiment, an objective function evaluation, or simply a function evaluation. We will use the term `iteration' to refer to a fixed number of function evaluations (usually one) performed by a simulation optimization algorithm.\\

A note of caution while using SO methods is to incorporate as much domain specific knowledge as possible in the use of an SO algorithm. This may be in terms of (1) screening relevant input variables, (2) scaling and range reduction of decision variables, (3) providing good initial guesses for the algorithm; and (4) gleaning information from known problem structure, such as derivative estimates.

Table \ref{tab:classification_optiproblems} classifies the techniques that are usually most suitable in practice for different scenarios in the universe of optimization problems. Certain broad classes of algorithms, such as random search methods, may be applicable to all of these types of problems, but they are often most suitable when dealing with pathological problems (e.g., problems with discontinuities, nonsmoothness) and are often used because they are relatively easy to implement.

\begin{table*}[h]
  \centering
  \caption{Terminology of optimization problems}
  \begin{tabular}[h]{l|p{2in}|p{1in}}
    \toprule
    & Algebraic model available & Unknown/complex problem structure \\
    \midrule
    Deterministic & Traditional math programming (linear, integer, and nonlinear programming)  & Derivative-free \newline optimization \\
    Uncertainty present & Stochastic programming, robust optimization & Simulation \newline optimization\\
    \bottomrule
  \end{tabular}
  \label{tab:classification_optiproblems}
\end{table*}

The possibilities of combining simulation and optimization procedures are vast: simulation with optimization-based iterations; optimization with simulation-based iterations; sequential simulation and optimization; and alternate simulation and optimization are four such paradigms. A recent paper by Figueira and Almada-Lobo \cite{fa14} delves into the taxonomy of such problems, and provides a guide to choosing an appropriate approach for a given problem. As detailed by Meketon \cite{meketon87}, different techniques may be applicable or more suitable depending on how much is known about the underlying simulation, such as its structure or associated probability distributions. We focus on approaches that are applicable in situations where  all the optimization scheme has to work with are evaluations of $f(x,y,\omega)$ and $g(x,y,\omega)$, or simply, observations with noise.

\subsection{Relationship to other fields}
\label{sec:relationship}

\paragraph{Mathematical Programming.} As mentioned earlier, most mathematical programming methods rely on the presence of an algebraic model. The availability of an algebraic model has many obvious implications to a mathematical programming expert, including the ability to evaluate a function quickly, the availability of derivative information, and the possibility of formulating a dual problem. None of these may be possible to do/obtain in an SO setting.

In the case with continuous decisions, derivative information is often hard to estimate accurately through finite differences, either due to the stochastic noise associated with objective function evaluations, or due to the large expense associated with obtaining function evaluations, or both. The inherent stochasticity in output also renders automatic differentiation (AD) \cite{rall81,gw08} tools not directly applicable. Moreover, automatic differentiation may not be used when one has no access to source code, does not possess an AD interface to proprietary simulation software, and, of course, when one is dealing with a physical experiment. The lack of availability of derivative information has further implications---it complicates the search for descent directions, proofs of convergence, and the characterization of optimal points.

Simulation optimization, like stochastic programming, also attempts to optimize under uncertainty. However, stochastic programming differs in that it makes heavy use of the model structure itself \cite{bl11}. Optimization under uncertainty techniques that make heavy use of mathematical programming are reviewed in \cite{sahinidis04}.

\paragraph{Derivative-Free Optimization.} Both Simulation Optimization and Derivative-Free Optimization (DFO) are referred to in the literature as black-box optimization methods. Output variability is the key factor that distinguishes SO from DFO, where the output from the simulation is deterministic. However, there are many approaches to DFO that have analogs in SO as well (e.g., response surfaces, direct search methods, metaheuristics), cf. Section \ref{sec:algos}.

Another distinction is that most algorithms in DFO are specifically designed keeping in mind that function evaluations or simulations are expensive. This is not necessarily the case with SO algorithms.

With regard to rates of convergence, SO algorithms are generally inefficient and convergence rates are typically very slow. In general, one would expect SO to have a slower convergence rate than DFO algorithms simply because of the additional complication of uncertainty in function evaluations. As explained in \cite{csv09}, some DFO algorithms, under certain assumptions, expect rates that are closer to linear than quadratic, and therefore early termination may be suitable. As described in some detail by \cite{fu94}, the best possible convergence rates for SO algorithms are generally $\mathcal{O}(1/\sqrt k)$, where $k$ is the number of samples. This is true from the central limit theorem that tells us the rate at which the best possible estimator converges to the true expected function value at a point. This implies that though one would ideally incorporate rigorous termination criteria in algorithm implementations, most practical applications have a fixed simulation or function evaluation budget that is reached first. \\

\paragraph{Machine Learning.} Several subcommunities in the machine learning community address problems closely related to simulation optimization. Traditional machine learning settings assume the availability of a fixed dataset. Active learning methods \cite{cgj96,settles10} extend machine learning algorithms to the case where the algorithms are allowed to query an oracle for additional data to infer better statistical models. Active learning is closely related in that this choice of sampling occurs at every iteration in a simulation optimization setting as well. The focus of active learning is usually to learn better predictive models rather than to perform optimization.

Reinforcement learning \cite{sb98} is broadly concerned with what set of actions to take in an environment to maximize some notion of cumulative reward. Reinforcement learning methods have strong connections to information theory, optimal control, and statistics. The similarity with simulation optimization is that the common problem of exploration of the search space vs. exploitation of known structure of the cost function arises. However, in the reinforcement learning setting, each action usually also incurs a cost, and the task is to maximize the accumulated rewards from all actions---as opposed to finding a good point in the parameter space eventually.

Policy gradient methods \cite{pvs03}  are a subfield of reinforcement learning, where the set of all possible sequences of actions form the policy space, and a gradient in this policy space is estimated and a gradient ascent-type method is then used to move to a local optimum. Bandit optimization \cite{gittins89} is another subfield of reinforcement learning that involves methods for the solution to the multi-armed bandit problem. The canonical example involves a certain number of slot machines, and a certain total budget to play them. Here, each choice of sample corresponds to which slot machine to play. Each play on a slot machine results in random winnings. This setting is analogous to discrete simulation optimization (DOvS) over finite sets, although with a different objective \cite{rp12}. Again, in DOvS over finite sets, we are only concerned with finding the best alternative eventually, whereas the cumulative winnings is the concern in the multi-armed bandit problem.

\paragraph{Relationship to other fields.}

Most, if not all, simulation optimization procedures have elements that are derived from or highly related to several other fields. Direct search procedures and response surface methodologies (RSM) have strong relationships with the field of experimental design. RSM, sample path optimization procedures, and gradient-based methods heavily incorporate ideas from mathematical programming. RSM also involves the use of nonparametric and Bayesian regression techniques, whereas estimation of distribution algorithms involves probabilistic inference, and therefore these techniques are related to statistics and machine learning. Simulation optimization has been described as being part of a larger field called computational stochastic optimization. More information is available at \cite{cso}.

\section{Applications}
\label{sec:applications}
SO techniques are most commonly applied to either (1) discrete-event simulations, or (2) systems of stochastic nonlinear and/or differential equations.

As mentioned in \cite{fu94}, discrete event simulations can be used to model many real-world systems such as queues, operations, and networks. Here, the simulation of a system usually involves switching or jumping from one state to another at discrete points in time as events occur. The occurrence of events is modeled using probability distributions to model the randomness involved.

Stochastic differential equations may be used to model phenomena ranging from financial risk \cite{merton74} to the control of nonlinear systems \cite{sg95} to the electrophoretic separation of DNA molecules \cite{cd10}.

With both discrete-event simulations and stochastic differential equation systems, there may be several parameters that one controls that affect some performance measure of the system under consideration, which are essentially degrees of freedom that may be optimized through SO techniques. Several applications of SO from diverse areas have been addressed in the literature and we list some of them in Table \ref{tab:applications}.

\begin{table*}[htbp]
  \centering
  \caption{Partial list of published works that apply simulation optimization}
    \begin{tabular}{p{2in}p{4in}}
    \toprule
    Domain of application & Application and citations \\
    \midrule
    Operations & Buffer location \cite{lds98}, nurse scheduling \cite{tr10}, inventory management \cite{kn05,swr06}, health care \cite{dfi03}, queuing networks \cite{fh97a,bhatnagar05,mbh07} \\
    Manufacturing & PCB production \cite{da00a}, engine manufacturing \cite{sl12}, production planning \cite{kb01,kleijnen93}, manufacturing-cell design \cite{iwt01}, kanban sizing \cite{hbu96} \\
   Medicine and biology & Protein engineering \cite{rka12}, cardiovascular surgery \cite{xfsme12}, breast cancer epidemiology \cite{fdfk05}, bioprocess control \cite{wrbr01,rw03}, ECG analysis \cite{gkvh02}, medical image analysis \cite{m+07} \\
   Engineering & Welded beam design \cite{yd10}, solid waste management \cite{yeomans07}, pollution source identification \cite{ayvaz10}, chemical supply chains \cite{j+04}, antenna design \cite{p+08}, aerodynamic design \cite{xd02,xd05,xd05a,kr05}, distillation column optimization \cite{r+01}, well placement \cite{b+05}, servo system control \cite{rppp11}, power systems \cite{e+07}, radar analysis \cite{k+06}\\
   Computer science, networks, electronics & Server assignment \cite{kk10}, wireless sensor networks \cite{dsa11}, circuit design \cite{li09}, network reliability \cite{khn07} \\
   Transportation and logistics & Traffic control and simulation \cite{yp06,b+07,ob10}, metro/transit travel times \cite{hf95,yb09}, air traffic control \cite{khi97,hh01} \\
    \bottomrule
    \end{tabular}%
  \label{tab:applications}%
\end{table*}%

\section{Algorithms}
\label{sec:algos}

Algorithms for SO are diverse, and their applicability may be highly dependent on the particular application. For instance, algorithms may (1) attempt to find local or global solutions; (2) address discrete or continuous variables; (3) incorporate random elements or not; (4) be tailored for cases where function evaluations are expensive; (5) emphasize exploration or exploitation to different extents; (6) assume that the uncertainty in simulation output is homoscedastic or that it comes from a certain probability distribution; or (7) rely on underlying continuity or differentiability of the expectation (or some function of a chosen moment) of the simulation output. The sheer diversity of these algorithms also makes it somewhat difficult to assert which one is better than another in general, and also makes it hard to compare between algorithms or their implementations.

As mentioned in Section \ref{sec:relationship}, many algorithms that are available for continuous simulation optimization have analogs in derivative-based optimization and in derivative-free optimization, where function evaluations are deterministic. In any case, the key lies in the statistics of how noise is handled, and how it is integrated into the optimization scheme. We will provide pointers to references that are applicable to simulation optimization in particular. A comprehensive review of methods for derivative-free optimization is available in \cite{rs13}.

Each major subsection below is accompanied by pseudocode to give researchers and practitioners unfamiliar with the field an idea of the general approach taken by each of these algorithms. Many of the sections include pointers to convergence proofs for individual algorithms. Optimality in simulation optimization is harder to establish than in mathematical programming or derivative-free optimization due to the presence of output variability. Notions of optimality for simulation optimization are explored in \cite{fu94}; for the discrete case, \cite{xnh10}, for instance, establishes conditions for local convergence, where a point being `better' than its $2m+1$ neighboring solutions is said to be locally optimal. There has also been some work in establishing Karush-Kuhn-Tucker (KKT) optimality conditions for multiresponse simulation optimization \cite{bck09}. Globally convergent algorithms will locate the global optimal solution eventually, but assuring this would require all feasible solutions to be evaluated through infinite observations; in practice, a convergence property that translates to a practical stopping criterion may make more sense \cite{hn09}.

Based on their scope, the broad classes of algorithms are classified in Table~\ref{tab:classify}. Algorithms are classified based on whether they are applicable to problems with discrete/continuous variables, and whether they focus on global or local optimization. However, there may be specific algorithms that have been tweaked to make them applicable to a different class as well, which may not be captured by this table.

\begin{table*}[htbp]
  \caption{Classification of simulation optimization algorithms}
  \begin{tabular}[htbp]{l|cc|cc}
    \toprule
    Algorithm class & Discrete & Continuous & Local & Global \\
    \midrule
    Ranking and selection & $\times$  & & & $\times$\\
    Metaheuristics & $\times$ & $\times$ & & $\times$\\
    Response surface methodology & & $\times$ & $\times$ & $\times$ \\
    Gradient-based methods & & $\times$  & $\times$  & \\
    Direct search & $\times$ & $\times$  & $\times$ & \\
    Model-based methods &  $\times$ & $\times$ & $\times$ & $\times$ \\
    Lipschitzian optimization & & $\times$ & & $\times$ \\
    \bottomrule
  \end{tabular}
  \label{tab:classify}
\end{table*}

\subsection{Discrete optimization via simulation}

Discrete optimization via simulation is concerned with finding optimal settings for variables that can only take discrete values. This may be in the form of \textit{integer-ordered} variables or \textit{categorical} variables \cite{ph11}. Integer-ordered variables are allowed to take on integer or discrete values within a finite interval, where the order of these values translates to some physical interpretation. For example, this could be the number of trucks available for vehicle routing, or the set of standard pipe diameters that are available for the construction of a manufacturing plant. Categorical variables refer to more general kinds of discrete decisions, ranging from conventional on-off (0-1 or binary) variables to more abstract decisions such as the sequence of actions to take given a finite set of actions. It should be noted that though integer-ordered variables, for instance, may be logically represented using binary variables, it may be beneficial to retain them as integer-ordered to exploit correlations in objective function values between adjacent integer values.

A rich literature in DOvS has developed over the last 50 years, and the specific methods developed are tailored to the specific problem setting. Broadly, methods are tailored for finite or for very large/potentially infinite parameter spaces.

\subsubsection{Finite parameter spaces}

In the finite case, where the number of alternatives is small and fixed, the primary goal is to decide how to allocate the simulation runs among the alternatives. In this setting, there is no emphasis on `search', as the candidate solution pool is small and known; each iteration is used to infer the best, in some statistical sense, simulation run(s) to be performed subsequently.

The optimization that is desired may differ depending on the situation, and could involve:
\begin{enumerate}
\item The selection of the best candidate solution from a finite set of alternatives;
\item The comparison of simulation performance measures of each alternative to a known standard or control; or
\item The pairwise comparison between all solution candidates.
\end{enumerate}

Item (1) is referred to as the \textit{ranking and selection} problem. Items (2) and (3) are addressed under literature on \textit{multiple comparison procedures}, with the former referred to as \textit{multiple comparisons with a control}.

\paragraph{Ranking and Selection.} In traditional ranking and selection, the task is to minimize the number of simulation replications while ensuring a certain probability of correct selection of alternatives. Most procedures try to guarantee that the design ultimately selected is better than all competing alternatives by $\delta$ with a probability at least $1-\alpha$. $\delta$ is called the indifference zone, and is the value deemed to be sufficient to distinguish between expected performance among solution candidates.

Conventional procedures make use of the Bonferroni inequality which relates probabilities of the occurrence of multiple events with probabilities of each event. Other approaches involve the incorporation of covariance induced by, for example, the use of common random numbers to expedite the algorithmic performance over the more conservative Bonferroni approach. Kim and Nelson~\cite{kn06,kn07} and Chick~\cite{chick06} provide a detailed review and provide algorithms and procedures for this setting. Extensions of fully sequential ranking and selection procedures to the constrained case have been explored as well, e.g., \cite{ak10}.

An alternative formulation of the ranking and selection of the problem would be to try to do the best within a specified computational budget, called the \textit{optimal computing budget allocation} formulation \cite{chen95}. Chen et al. \cite{cydc09} present more recent work, while the stochastically constrained case is considered in \cite{l+12}.

Recent work \cite{hp13} in the area of DOvS over finite sets provides a quick overview of the field of ranking and selection, and considers general probability distributions and the presence of stochastic constraints simultaneously.

A basic ranking and selection procedure \cite{kn07} is outlined in Algorithm~\ref{alg:ranking}, where it is assumed that independent data comes from normal distributions with unknown, different variances.

\begin{algorithm}
  \caption{Basic ranking and selection procedure for SO}
  \label{alg:ranking}
  \begin{algorithmic}[1]
    \Require Confidence level $1-\alpha$, indifference zone parameter $\delta$
    \State Take $n_0$ samples from each of the $1, \ldots, K$ potential designs
    \State Compute sample means, $\bar{t}_{k,n_0}$ and sample variances, $S_k$, for each of the designs
    \State Determine how many new samples, $N_k:= \max\left\{n_0, \left\lceil \frac{\psi^2S_k^2}{\delta^2} \right\rceil \right\}$, to take from each system, where the Rinott constant $\psi$ is obtained from \cite{bsg95}
    \State Select the system with the best new sample mean, $\bar{t}_{k,N_k+n_0}$.
  \end{algorithmic}
\end{algorithm}

\paragraph{Multiple comparison procedures.} Here, a number of simulation replications are performed on all the potential designs, and conclusions are made by constructing confidence intervals on the performance metric. The main ideas and techniques for multiple comparisons in the context of pairwise comparisons, or against a known standard are presented in \cite{ht87}, \cite{fu94}, and \cite{hsu96}. Recent work in multiple comparisons with a control include \cite{kim05} and \cite{ng01}, which provide fully sequential and two-stage frequentist procedures respectively; and \cite{xf13}, which addresses the problem using a Bayesian approach.\\

Comprehensive treatment of ranking and selection and multiple comparison procedures may be found in Goldsman and Nelson \cite{gn98} and Bechhofer et al. \cite{bsg95}. A detailed survey that traces the development of techniques in simulation optimization over finite sets is available in \cite{ts04}.

\subsubsection{Large/Infinite parameter spaces}

To address DOvS problems with a large number of potential alternatives, algorithms that have a search component are required. Many of the algorithms that are applicable to the continuous optimization via simulation case are, with suitable modifications, applicable to the case with large/infinite parameter spaces. These include (1) ordinal optimization (2) random search methods and (3) direct search methods.

Ordinal optimization methods \cite{ho99} are suitable when the number of alternatives is too large to find the globally optimal design in the discrete-event simulation context. Instead, the task is to find a satisfactory solution with some guarantees on quality (called alignment probability) \cite{lh97}. Here, the focus is on sampling a chosen subset of the solutions and evaluating them to determine the best among them. The key lies in choosing this subset such that it contains a subset of satisfactory solutions. The quality or satisfaction level of this selected subset can be quantified \cite{chen96}. A comparison of subset selection rules is presented in \cite{jhz06} and the multi-objective case is treated in \cite{tlc07}.

Random search methods include techniques such as simulated annealing (e.g., \cite{aa99}), genetic algorithms, stochastic ruler methods (e.g., \cite{ym92}), stochastic comparison (e.g., \cite{ghz99}), nested partitions (e.g., \cite{so00}), ant colony optimization (e.g., \cite{ds04,db05}), and tabu search (e.g., \cite{gh02}). Some of these---simulated annealing, genetic algorithms, and tabu search---are described in Section~\ref{sec:metaheuristics}). Ant colony optimization is described under model-based methods (cf. Section~\ref{sec:aco}). Proofs of global convergence, i.e., convergence to a global solution, or local convergence are available for most of these algorithms \cite{hn09} (note that these definitions differ from mathematical programming where \textit{global convergence} properties ensure convergence to a \textit{local optimum} regardless of the starting point).

Nested partition methods \cite{so07} attempt to adaptively sample from the feasible region. The feasible region is then partitioned, and sampling is concentrated in regions adjudged to be the most promising by the algorithm from a pre-determined collection of nested sets. Hong and Nelson propose the COMPASS algorithm \cite{hn06a} which uses a unique neighborhood structure, defined as the most promising region that is fully adaptive rather than pre-determined; a most promising `index' is defined that classifies each candidate solution based on a nearest neighbor metric. More recently, the Adaptive Hyberbox Algorithm \cite{xnh13} claims to have superior performance on high-dimensional problems (problems with more than ten or fifteen variables); and the R-SPLINE algorithm \cite{wps12}, which alternates between a continuous search on a continuous piecewise-linear interpolation and a discrete neighborhood search, compares favorably as well.

A review of random search methods is presented in \cite{andradottir06,o06}. Recent progress, outlines of basic algorithms, and pointers to specific references for some of these methods are presented in \cite{b+09}, \cite{hn09}, and \cite{nelson10}.

Direct search methods such as pattern search and Nelder-Mead simplex methods are elaborated on in Section~\ref{sec:directsearch}.

\subsection{Response surface methodology}

Response surface methodology (RSM) is typically useful in the context of continuous optimization problems and focuses on learning input-output relationships to approximate the underlying simulation by a surface (also known as a metamodel or surrogate model) for which we define a functional form. This functional form can then be made use of by leveraging powerful derivative-based optimization techniques. The literature in RSM is vast and equivalent approaches have variously been referred to as multi-disciplinary design optimization, metamodel-based optimization, and sequential parameter optimization. RSM was originally developed in the context of experimental design for physical processes \cite{bw51}, but has since been applied to computer experiments. Metamodel-based optimization is a currently popular technique for addressing simulation optimization problems \cite{bm06,kleijnen08}.

\begin{algorithm}
  \caption{Basic RSM procedure}
  \label{alg:rsm}
  \begin{algorithmic}[1]
    \Require Initial region of approximation $\mathcal{X}$, choice of regression surface $r$
    \While{under simulation budget and not converged}
    \State Perform a design of experiments in relevant region, using $k$ data points
    \State $t_i \gets$ simulate($x_i),\quad i=\{1,\ldots,k\}$ \qquad \{Evaluate noisy function $f(x_i,\omega)$\}
    \State $\lambda^* \gets \arg\min_\lambda \sum (t_i - r(x_i, \lambda))^2 $ \qquad \{Fit regression surface $r$ through points using squared loss function\}
    \State $x^* \gets \{\arg\min_x r(x,\lambda^*): x \in \mathcal{X}\}$ \qquad \{Optimize surface\}
    \State Update set of available data points and region of approximation
    \EndWhile
  \end{algorithmic}
\end{algorithm}

Different response surface algorithms differ in the choice between regression and interpolation; the nature of the functional form used for approximation (polynomials, splines, Kriging, radial basis functions, neural networks); the choice of how many and where new samples must be taken; and how they update the response surface.

RSM approaches can either (1) build surrogate models that are effective in local regions, and sequentially use these models to guide the search, or; (2) build surrogate models for the entire parameter space from space-filling designs, and then use them to choose samples in areas of interest, i.e., where the likelihood of finding better solutions is good according to a specified metric. A generic framework for RSM is presented in Algorithm \ref{alg:rsm}.

\paragraph{Classical sequential RSM.} Originally, RSM consisted of a Phase I, where first-order models were built using samples from a design of experiments. A steepest descent rule was used to move in a certain direction, and this would continue iteratively until the estimated gradient would be close to zero. Then, a Phase II procedure that built a more detailed quadratic model would be used for verifying the optimality of the experimental design. A thorough introduction to response surface methodology is available in \cite{mma09}. Recent work in the field includes automating RSM \cite{nopd00,nd09} and the capability to handle stochastic constraints \cite{akh09}.

\paragraph{Bayesian global optimization.} These methods seek to build a global response surface, commonly using techniques such as Kriging/Gaussian process regression \cite{ssw89,rw06}. Subsequent samples chosen based on some sort of improvement metric may balance exploitation and exploration. The seminal paper by Jones et al. \cite{jsw98} which introduced the EGO algorithm for simulations with deterministic output, uses Kriging to interpolate between function values, and chooses future samples based on an expected improvement metric \cite{mtz78ch2}. Examples of analogs to this for simulation optimization are provided in \cite{hanz06,kbn12}. The use of Kriging for simulation metamodeling is explored in \cite{vk04,kv05,kleijnen09}. Other criteria that have been used to choose samples are most probable improvement \cite{mockus89}, knowledge gradient for continuous parameters \cite{sfp11}, and maximum information gain \cite{nkks12}.

\paragraph{Trust region methods.} Trust region methods \cite{cgt00} can be used to implement sequential RSM. Trust regions provide a means of controlling the region of approximation, providing update criteria for surrogate models, and are useful in analyzing convergence properties. Once a metamodel or response surface, $g$, is built around a trust region center $x_i$, trust region algorithms involve the solution of the trust-region subproblem ($\min_s g(x_i+s): s \in \mathcal{B}(x_i,\Delta)$), where $\mathcal{B}$ is a ball defined by the center-radius pair $(x_i, \Delta)$. There are well-defined criteria to update the trust region center and radius \cite{cgt00} that will define the subsequent region of approximation.

The use of trust regions in simulation optimization is relatively recent, and has been investigated to some extent \cite{df06,chw13}. Trust-region algorithms have been used, for example, to optimize simulations of urban traffic networks \cite{ob10}.\\

\subsection{Gradient-based methods}

Stochastic approximation methods or gradient-based approaches are those that attempt to descend using estimated gradient information. Stochastic approximation techniques are one of the oldest methods for simulation optimization. Robbins and Monro \cite{rm51} and Kiefer and Wolfowitz \cite{kw52} were the first to develop stochastic approximation schemes in the early 1950s. These procedures initially were meant to be used under very restrictive conditions, but much progress has been made since then.

These methods can be thought of being analogous to steepest descent methods in derivative-based optimization. One may obtain direct gradients or may estimate gradients using some finite difference scheme. Direct gradients may be calculated by a number of methods: (1) Perturbation Analysis (specifically, Infinitesimal Perturbation Analysis) (PA or IPA), (2) Likelihood Ratio/Score Function (LR/SF), and (3) Frequency Domain Analysis (FDA). Detailed books on these methods are available in the literature \cite{hc91,glasserman91,rs93,pflug96,fh97} and more high-level descriptions are available in papers \cite{ts04,fu02}. Most of these direct methods, however, are either applicable to specific kinds of problems, need some information about underlying distributions, or are difficult to apply. Fu \cite{fu02} outlines which methods are applicable in which situations, and Tekin and Sabuncuoglu \cite{ts04} discuss a number of applications that have used these methods.

Stochastic approximation schemes attempt to estimate a gradient by means of finite differences. Typically, a forward difference estimate would involve sampling at least $n + 1$ distinct points, but superior performance has been observed by simultaneous perturbation estimates that require samples at just two points \cite{spall03ch07}, a method referred to as Simultaneous Perturbation Stochastic Approximation (SPSA). The advantage gained in SPSA is that the samples required are now independent of the problem size, and, interestingly, this has been shown to have the same asymptotic convergence rate as the naive method that requires $n + 1$ points \cite{spall92}. A typical gradient-based scheme is outlined in Algorithm \ref{alg:stapp}.

\begin{algorithm}
  \caption{Basic gradient-based procedure}
  \label{alg:stapp}
  \begin{algorithmic}[1]
    \Require Specify initial point, $x_0$. Define initial parameters such as step size ($\alpha$), distances between points for performing finite difference, etc.
    \State $i \gets 0$
    \While{under simulation budget and not converged}
    \State Perform required simulations, $t_{i}^{j_i} \gets \text{simulate}(x_{i})$, with $j_i$ replications to estimate gradient, $\widehat{J}$, using either IPA, LR/SF, FDA or finite differences
    \State $x_{i+1} \gets x_i - \alpha\widehat{J}$
    \State $i \gets i+1$
    \EndWhile
  \end{algorithmic}
\end{algorithm}

Recent extensions of the SPSA method include introducing a global search component to the algorithm by injecting Monte Carlo noise during the update step \cite{mc08}, and using it to solve combined discrete/continuous optimization problems \cite{ws11}. Recent work also addresses improving Jacobian as well as Hessian estimates in the context of the SPSA algorithm \cite{spall09}.  Much of the progress in stochastic approximation has been cataloged in the proceedings of the Winter Simulation Conference over the years (\texttt{http://informs-sim.org/}). A recent review of stochastic approximation methods is available in \cite{spall12ch07}, and an excellent tutorial and review of results in stochastic approximation is presented in \cite{pk11}.

\subsection{Sample path optimization}

Sample path optimization involves working with an estimate of the underlying unknown function, as opposed to the function itself. The estimate is usually a consistent estimator such as the sample mean of independent function evaluations at a point, or replications. For instance, one may work with $F_n = \frac{1}{n}\sum_{i=1}^n f(x,y,\omega_i)$, instead of the underlying function ${E}[f(x,y,\omega)]$ itself. It should be noted that the functional form of $F_n$ is still unknown, it is just that $F_n$ can be observed or evaluated at a point in the search space visited by an algorithm iteration. The alternative name of sample average approximation reflects this use of an estimator.

As the algorithm now has to work with an estimator, a deterministic realization of the underlying stochastic function, sophisticated techniques from traditional mathematical programming can now be leveraged. Sample path methods can be viewed as the use of deterministic optimization techniques within a well-defined stochastic setting. Yet another name for them is stochastic counterpart. Some of the first papers using sample path methods are \cite{hs91} and \cite{shapiro91}. Several papers \cite{rs93,cs94,gor94,shapiro96,df06} discuss convergence results and algorithms in this context.

\subsection{Direct search methods}
\label{sec:directsearch}

Direct search can be defined as the sequential examination of trial solutions generated by a certain strategy \cite{hj61}. As opposed to stochastic approximation, direct search methods rely on direct comparison of function values without attempting to approximate derivatives. Direct search methods typically rely on some sort of ranking of quality of points, rather than on function values.

Most direct search algorithms developed for simulation optimization are extensions of ideas for derivative-free optimization. A comprehensive review of classical and modern methods is provided in \cite{klt03}. A formal theory of direct search methods for stochastic optimization is developed in \cite{trosset00}. Direct search methods can be tailored for both discrete and continuous optimization settings. Pattern search and Nelder-Mead simplex procedures are the most popular direct search methods. There is some classical as well as relatively recent work done on investigating both pattern search methods \cite{trosset00,af01,ls02} and Nelder-Mead simplex algorithms \cite{nm65,bi96,hw00,chang12} and their convergence in the context of simulation optimization.

These methods remain attractive as they are relatively easy to describe and implement, and are not affected if a gradient does not exist everywhere, as they do not rely on gradient information. Since conventional procedures can be affected by noise, effective sampling schemes to control the noise are required. A basic Nelder-Mead procedure is outlined in Algorithm~\ref{alg:nelder}.

\begin{algorithm}
  \caption{Basic Nelder-Mead simplex procedure for SO}
  \label{alg:nelder}
  \begin{algorithmic}[1]
    \Require A set of $n-1$ points in the parameter space to form the initial simplex
    \While{under simulation budget and not converged}
    \State Generate a new candidate solution, $x_i$, through simplex centroid reflections, contractions or other means
    \State $t_i^{j_i} \gets$ simulate($x_{i}$), $\quad i = \{i-n+1, \ldots, i\}, j_i=\{1,\ldots,N_i\}$ \qquad \{Evaluate noisy function $f(x,\omega)$ $N_i$ times, where $N_i$ is determined by some sampling scheme\}
    \State Calculate $\frac{\sum_{j_i} t_i^{j_i}}{N_i}$, or some similar metric to determine which point (i.e., with the highest metric value) should be eliminated
    \EndWhile
  \end{algorithmic}
\end{algorithm}

\subsection{Random search methods}
\label{sec:metaheuristics}

\subsubsection{Genetic algorithms}

Genetic algorithms use concepts of mutation and selection \cite{reeves97,whitley94}. In general, a genetic algorithm works by creating a population of strings and each of these strings are called chromosomes. Each of these chromosome strings is basically a vector of point in the search space. New chromosomes are created by using selection, mutation and crossover functions. The selection process is guided by evaluating the fitness (or objective function) of each chromosome and selecting the chromosomes according to their fitness values (using methods such as mapping onto Roulette Wheel). Additional chromosomes are then generated using crossover and mutation functions. The cross over and mutation functions ensure that a diversity of solutions is maintained. Genetic algorithms are popular as they are easy to implement and are used in several commercial simulation optimization software packages (Table~\ref{tab:comm_packages}). The GECCO (Genetic and Evolutionary Computation Conference) catalogs progress in genetic algorithms and implementations.

\subsubsection{Simulated annealing}

Simulated Annealing uses a probabilistic method that is derived from the annealing process in which the material is slowly cooled so that, while its structure freezes, it reaches a minimum energy state \cite{kgv83,bt93}. Starting with a current point $i$ in a state $j$, a neighborhood point $i'$ of the point $i$ is generated. The algorithm moves from point $i$ to $i'$ using a probabilistic criteria that is dependent on the `temperature' in state $j$. This temperature is analogous to that in physical annealing, and serves here as a control parameter. If the solution at $i'$ is better than the existing solution, then this new point is accepted. If the new solution is worse than existing solution, then the probability of accepting the point is defined as $\exp(-(f(i')- f(i))/T(j))$, where $f(.)$ is the value of objective function at a given point, and $T(j)$ is temperature at the state $j$.  After a certain number of neighborhood points are evaluated, the temperature is decreased and new state is $j+1$ is created. Due to the exponential form, the probability of acceptance of a neighborhood point is higher at high temperature, and is lower as temperature is reduced. In this way, the algorithm searches for a large number of neighborhood points in the beginning, but a lower number of points as temperature is reduced.

Implementation of simulated annealing procedures require choosing parameters such as the initial and final temperatures, the rate of cooling, and number of function evaluations at each temperature. A variety of cooling `schedules' have been suggested in \cite{ceg88} and \cite{hajek88}. Though simulated annealing was originally meant for optimizing deterministic functions, the framework has been extended to the case of stochastic simulations \cite{aat99}. The ease of implementing a simulated annealing procedure is high and it remains a popular technique used by several commercial simulation optimization packages.

\subsubsection{Tabu search}

Tabu search \cite{glover90} uses special memory structures (short-term and long-term) during the search process that allow the method to go beyond local optimality to explore promising regions of the search space. The basic form of tabu search consists of a modified neighborhood search procedure that employs adaptive memory to keep track of relevant solution history, together with strategies for exploiting this memory \cite{gp10}. More advanced forms of tabu search and its applications are described in \cite{gl97}.


\subsubsection{Scatter search}

Scatter search and its generalized form, path relinking, were originally introduced by Glover and Laguna \cite{gl00}. Scatter search differs from other evolutionary approaches (such as Genetic Algorithms (GA)) by using strategic designs and search path construction from a population of solutions as compared to randomization (by crossover and mutation in GA). Similar to Tabu search, Scatter Search also utilize adaptive memory in storing best solutions \cite{gl00,mlg06}. Algorithm~\ref{alg:scatter} provides the scatter search algorithm.

 \begin{algorithm}
   \caption{Basic scatter search procedure for SO}
   \label{alg:scatter}
   \begin{algorithmic}[1]
     \Require An initial set of trial points $x \in P$, chosen to be diversified according to a pre-specified metric
     \State $t_j \gets$ simulate($x_j$), where $j = 1, \ldots, |P|$
     \State $k \gets 0$
     \State Use a comparison procedure (such as ranking and selection) to gather the best $b$ solutions (based on objective value or diversity) from the current set of solutions $P$, called the reference set, $R_k$
     \State $R_{-1} = \varnothing$
     \While{under simulation budget and $R_k \neq R_{k-1}$}
     \State $k \gets k + 1$
     \State Choose $S_i \subset R$, where $i = 1, \ldots, r$ \{Use a subset generation procedure to select $r$ subsets of set $R$, to be used as a basis for generating new solution points\}
    \For{$i = 1$ to $r$}
     \State Combine the points in $S_i$, to form new solution points, $x_j$, where $j \in \mathcal{J} = |P|+1, \ldots, |P|+J$, using weighted linear combinations, for example
     \State $t_j \gets$ simulate($x_j$), $j \in \mathcal{J}$ \{sample the objective function at new trial solutions\}
     \State Update sets $R_k$, $P$
    \EndFor
     \EndWhile
   \end{algorithmic}
 \end{algorithm}

\subsection{Model-based methods}

 Model-based simulation optimization methods attempt to build a probability distribution over the space of solutions and use it to guide the search process.

\subsubsection{Estimation of distribution algorithms}

Estimation of distribution algorithms (EDAs) \cite{ll02} are model-based methods that belong to the evolutionary computation field. However, generation of new candidate solutions is done by sampling from the inferred probability distribution over the space of solutions, rather than, say, a genetic operator such as crossover or mutation. A comprehensive review of estimation of distribution algorithms is presented in \cite{fhm06}. EDAs usually consider interactions between the problem variables and exploit them through different probability models.

Cross-entropy methods and Model Reference Adaptive Search (MRAS) are discussed next and can be seen as specific instances of EDAs.

\paragraph{Cross-Entropy Methods.} Cross-entropy methods first sample randomly from a chosen probability distribution over the space of decision variables. For each sample, which is a vector defining a point in decision space, a corresponding function evaluation is obtained. Based on the function values observed, a pre-defined percentile of the best samples are picked. A new distribution is built around this `elite set' of points via maximum likelihood estimation or some other fitting method, and the process is repeated. One possible method that implements cross-entropy is formally described in Algorithm \ref{alg:crossentropy}.

\begin{algorithm}
  \caption{Pseudocode for a simple cross-entropy implementation}
  \label{alg:crossentropy}
  \begin{algorithmic}[1]
    \Require $\theta$, an initial set of parameters for a pre-chosen distribution $p(x; \theta)$ over the set of decision variables; $k$, a number of simulations to be performed; $e$, the number of elite samples representing the top $\delta$ percentile of the $k$ samples
    \While{under simulation budget and not converged}
    \For{$i=1 \to k$}
    \State sample $x_i$ from $p(x;\theta)$
    \State $t_i \gets$ simulate($x_i$)
    \EndFor
    \State $E \gets \emptyset$
    \For{$i = 1 \to e$}
    \State $E_j \gets \arg\max_{i \notin E} t_i$
    \EndFor
    \State $p(x;\theta) \gets$ fit($x_E$)
    \EndWhile
  \end{algorithmic}
\end{algorithm}

The method is guaranteed (probabilistically) to converge to a local optimum, but it also incorporates an exploration component as random samples are obtained at each step. However, the intuition behind the selection of subsequent samples can be shown to be analogous to minimizing the Kullback-Leibler divergence (KL-divergence) between the optimal importance sampling distribution and the distribution used in the current iterate \cite{rk04}.

There exist variants of the cross-entropy method to address both continuous \cite{kpr06} and discrete optimization \cite{rubinstein99} problems. A possible modification is to use mixtures of distributions from current and previous iterations, with the current distribution weighted higher. This can be done by linearly interpolating the mean covariance in the case of Gaussian distributions. This also helps in avoiding singular covariance matrices. Cross-entropy can also deal with noisy function evaluations, with irrelevant decision variables, and constraints \cite{kpr06}. If decision variables are correlated, the covariance of the distribution will reflect this.

The immediately apparent merits of cross-entropy methods are that they are easy to implement, require few algorithmic parameters, are based on fundamental principles such as KL-divergence and maximum likelihood, and give consistently accurate results \cite{kpr06}. A potential drawback is that cross-entropy may require a significant number of new samples at every iteration. It is not clear as to how this would affect performance if samples were expensive to obtain. The cross-entropy method has analogs in  simulated annealing, genetic algorithms, and ant colony optimization, but differs from each of these in important ways \cite{dkmr05}.

More detailed information on the use of cross-entropy methods for optimization can be found in \cite{dkmr05}, a tutorial on cross-entropy and in \cite{rk04}, a monograph. The cross-entropy webpage, {http://iew3.technion.ac.il/CE/} provides up-to-date information on progress in the field.

\paragraph{Model reference adaptive search (MRAS).} The MRAS method \cite{hfm05,hfm07} is closely related to the cross-entropy method. It also works by minimizing the Kullback-Leibler divergence to update the parameters of the inferred probability distribution. However, the parameter update step involves the use of a sequence of implicit probability distributions. In other words, while the cross-entropy method uses the optimal importance sampling distribution for parameter updates, MRAS minimizes the KL-divergence with respect to the distribution in the current iteration, called the reference model.

\paragraph{Covariance Matrix Adaptation--Evolution Strategy (CMA-ES).} In the CMA-ES algorithm \cite{h06}, new samples are obtained from a multivariate normal distribution, and inter-variable dependencies are encoded in the covariance matrix. The CMA-ES method provides a way to update the covariance matrix. Updating the covariance matrix is analogous to learning an approximate inverse Hessian, as is used in Quasi-Newton methods in mathematical programming. The update of the mean and covariance is done by maximizing the likelihood of previously successful candidate solutions and search steps, respectively. This is in contrast to other EDAs and the cross-entropy method, where the covariance is updated by maximizing the likelihood of the successful points. Other sophistications such as step-size control, and weighting of candidate solutions are part of modern implementations \cite{cma-es}.

\subsubsection{Ant colony optimization}
\label{sec:aco}

Ant colony optimization methods \cite{ds04,db05} are heuristic methods that have been used for combinatorial optimization problems. Conceptually, they mimic the behavior of ants to find shortest paths between their colony and food sources. Ants deposit pheromones as they walk; and are more likely to choose paths with higher concentration of pheromones. This phenomenon is incorporated in a pheromone update rule, which increases the pheromone content in components of high-quality solutions, and causes evaporation of pheromones in less favorable regions. Probability distributions are used to make the transition between each iteration. These methods differ from EDAs in that they use an iterative construction of solutions.

This and other algorithms that incorporate self-organization in biological systems are said to use the concept of `swarm intelligence'. \\

\subsection{Lipschitzian optimization}
\label{sec:lipsch_opt}

Lipschitzian optimization is a class of space-partitioning algorithms for performing global optimization, where the Lipschitz constant is pre-specified. This enables the construction of global search algorithms with convergence guarantees. The caveat of having prior knowledge of the Lipschitz constant is overcome by the DIRECT (DIviding RECTangles) algorithm \cite{jps93} for deterministic continuous optimization problems. An adaptation of this for noisy problems is provided in \cite{df07}.

\section{Software}
\label{sec:software}

\subsection{Simulation optimization in commercial simulation software}

Many discrete-event simulation packages incorporate some methodology for performing optimization. A comprehensive listing of simulation software, the corresponding vendors, and the optimization packages and techniques they use can be found in Table \ref{tab:comm_packages}. More details on the specific optimization routines can be found in \cite{lk00}. OR/MS-Today, the online magazine of INFORMS, conducts a biennial survey of simulation software packages, the latest of which is available at \cite{orms2013survey}. The survey lists 43 simulation software packages, and 31 of these have some sort of optimization routine; fewer still have black-box optimizers that interact with the simulation.

\begin{table*}[htbp]
  \centering
  \caption{Simulation optimization packages in commercial simulation software}
    \begin{tabular}{p{1in}p{1.5in}p{1.5in}p{1.5in}}
    \toprule
    \textbf{Optimization package} & \textbf{Vendor} & \textbf{Simulation software supported} & \textbf{Optimization methodology} \\[15pt]
    \midrule
    AutoStat & Applied Materials, Inc. & AutoMod & Evolutionary strategy \\[15pt]
    Evolutionary Optimizer & Imagine That, Inc. & ExtendSim & Evolutionary strategy \\[15pt]
    OptQuest & OptTek Systems, Inc. & FlexSim, @RISK, Simul8, Simio, SIMPROCESS, AnyLogic, Arena,  Crystal Ball, Enterprise Dynamics, ModelRisk  &  Scatter search, tabu search, neural networks, integer programming \\[45pt]
    SimRunner & ProModel Corp. & ProModel, MedModel, ServiceModel & Genetic algorithms and evolutionary strategies \\[15pt]
    RISKOptimizer & Palisade Corp. &  @RISK & Genetic algorithm \\[15pt]
    WITNESS Optimizer & Lanner Group, Inc. & WITNESS & Simulated annealing, tabu search, hill climbing \\[5pt]
    GoldSim \newline Optimizer & GoldSim Technology Group & GoldSim & Box's complex method \\[15pt]
    Plant \newline Simulation \newline Optimizer & Siemens AG & Siemens PLM software      & Genetic algorithm \\[25pt]
    ChaStrobeGA &  N/A  & Stroboscope & Genetic algorithm \\[5pt]
    Global\newline Optimization toolbox & The MathWorks & SimEvents (Matlab) & Genetic algorithms, \newline simulated annealing,\newline pattern search \\
    \bottomrule
    \end{tabular}%
  \label{tab:comm_packages}%
\end{table*}%

\subsection{Academic implementations of simulation optimization}

Table~\ref{tab:SO_algos} contains a small subset of academic implementations of SO algorithms, and classifies them by type. Some of these are available for download from the web, some have code with suggested parameters in corresponding papers themselves, and others are available upon request from the authors.

\begin{table*}[h]
  \centering
  \caption{Academic simulation optimization implementations}
  \begin{tabular}{p{0.85in}lll}
    \toprule
    Algorithm & Type & Citation & Year\\
    \midrule
    Continuous & & & \\
    \midrule
      SPSA & Stochastic approximation & \cite{spall03ch07} & 1992\\ 
      SPSA 2nd Order & Stochastic approximation & \cite{spall03ch07} & 1999\\ 
      SKO & Global response surface &  \cite{hanz06} & 2006\\ 
      CE method & Cross-entropy & \cite{kpr06} & 2006 \\
      APS & Nested partitioning & \cite{ko07} & 2007 \\
      SNOBFIT & Multi-start local response surface & \cite{hn08} & 2008\\ 
      CMA-ES & Evolutionary strategy & \cite{cma-es} & 2011\\ 
      KGCP & Global response surface & \cite{sfp11} & 2011\\ 
      STRONG & Local response surface, trust region  & \cite{chw13} & 2011\\ 
      GR & Golden region search & \cite{ko11} & 2011 \\
      SNM & Direct search (Nelder-Mead) & \cite{chang12} & 2012\\
      DiceOptim & Global response surface & \cite{rgd12} & 2012\\
      \midrule
      Discrete & & & \\
      \midrule
      KG & Global response surface & \cite{fpd09} & 2009\\ 
      COMPASS & Neighborhood search (integer-ordered problems) & \cite{xnh10} & 2010 \\
      R-SPLINE & Neighborhood search (integer-ordered problems) & \cite{wps12} & 2012 \\
      \midrule
      Discrete and continuous & & &\\
      \midrule
      MRAS & Estimation of distribution & \cite{hfm05,hfm07} & 2005\\
      NOMADm & Mesh adaptive direct search & \cite{nomadm4.5} & 2007\\
      \bottomrule
  \end{tabular}
  \label{tab:SO_algos}
\end{table*}

\section{Comparison of algorithms}
\label{sec:comparisons}

As far as comparisons between algorithms are concerned, the literature does not yet provide a comprehensive survey of the performance of different implementations and approaches on large test beds. In this regard, simulation optimization lags behind other optimization fields such as linear, integer, and nonlinear programming, global optimization and even derivative-free optimization, where the first comprehensive comparison appeared in 2013 \cite{rs13}. A study of prior comparisons in simulation optimization is provided by \cite{ts04}, but these comparisons are fairly dated, are inconclusive about which algorithms perform better in different situations, and compare only a small subset of available algorithms. One difficulty lies in the inherent difficulty of comparing solutions between algorithms over true black-box simulations, as one does not usually know the true optimal point and can only compare between noisy estimates observed by the solvers. Less impeding difficulties, but difficulties nonetheless, include the need to interface algorithms to a common wrapper, the objective comparison with solvers that incorporate random elements as their results may not be reproducible, and lack of standard test simulations for purposes of benchmarking.

The benchmarking of algorithms in mathematical programming is usually done by performance profiles \cite{dm02}, where the graphs show the fraction of problems solved after a certain time. For derivative-free algorithms, data profiles are commonly used \cite{mw09}, where the fraction of problems solved after a certain number of iterations (function evaluations) or `simplex gradients' is shown. The definition of when a problem is `solved' may vary---when the true global optimum is known, the solutions found within a certain tolerance of this optimal value may be called solutions, but when this optimum is not known, the solvers that find the best solution (within a tolerance) for a problem, with respect to the other solvers being compared, may be said to have solved the problem. The latter metric may also be used when function evaluations are expensive, and no solver is able to reach within this tolerance given the limited simulation budget.

In both of these cases, the output of the simulations are deterministic, and so it is clear as to which algorithms have performed better than others on a particular problem. In simulation optimization, however, usually one does not know the true solution for the black box system, nor does one see deterministic output. All that one possesses are mean values and sample variances obtained from sample paths at different points. There does not exist a standard method to compare simulation optimization algorithms on large test beds. Many papers perform several macroreplications and report the macroreplicate average of the best sample means (along with the associated sample variance) at the end of the simulation budget. The issue with this is that the performance of the algorithms with different simulation budgets is not seen, as in the case of performance or data profiles. Other papers report the average number of evaluations taken to find a sample mean that is within the global tolerance for each problem. Here, results are listed for each problem and one does not get an idea of overall performance. In addition, the difference in sample variance estimates is not highlighted. As simulation optimization develops, there is also a need for methods of comparison of algorithms on test beds with statistically significant number of problems.

With regard to standardized simulation testbeds, to our knowledge, the only testbed that provides practical simulations for testing simulation optimization algorithms is available at www.simopt.org \cite{ph11}. At the point of writing this paper, just 20 continuous optimization problems were available from this repository. Most testing and comparisons happen with classical test problems in nonlinear optimization (many of which have been compiled in \cite{rs13} and available at http://archimedes.cheme.cmu.edu/?q=dfocomp), to which stochastic noise has been added. There is a need for more such repositories, not only for testing of algorithms over statistically significant sizes of problem sets, but for comparison between different classes of algorithms. The need for comparison is evident, given the sheer number of available approaches to solving simulation optimization problems, and the lack of clarity and lack of consensus on which types of algorithms are suitable in which contexts.

As observed by several papers \cite{f+00,ts04,hn09}, there continues to exist a significant gap between research and practice in terms of algorithmic approaches. Optimizers bundled with simulation software, as observed in Section~\ref{sec:software}, tend to make use of algorithms which seem to work well but do not come with provable statistical properties or guarantees of local or global convergence. Academic papers, on the other hand, emphasize methods that are more sophisticated and prove convergence properties. One reason that may contribute to this is that very few simulation optimization algorithms arising from the research community are easily accessible. We wholeheartedly encourage researchers to post their executable files, if not their source code. This could not only encourage practitioners to use these techniques in practice, but allow for comparisons between methods and the development of standardized interfaces between simulations and simulation optimization software.


\section{Conclusions}
\label{sec:conclusions}

The field of simulation optimization has progressed significantly in the last decade, with several new algorithms, implementations, and applications. Contributions to the field arise from researchers and practitioners in the industrial engineering/operations research, mathematical programming, statistics and machine learning, as well as the computer science communities. The use of simulation to model complex, dynamic, and stochastic systems has only increased with computing power and availability of a wide variety of simulation languages. This increased use is reflected in the identification and application of simulation and simulation optimization methods to diverse fields in science, engineering, and business. There also exist strong analogies between, and ideas that may be borrowed from recent progress in related fields. All of these factors, along with the ever increasing number of publications and rich literature in this area, clearly indicate the interest in the field of simulation optimization, and we have tried to capture this in this paper.

With increased growth and interest in the field, there come also opportunities. Potential directions for the field of simulation optimization are almost immediately apparent. Apart from the ability to handle simulation outputs from any well-defined probability distribution, the effective use of variance reduction techniques when possible, and the improvement in theory and algorithms, there is a requirement to address (1) large-scale problems with combined discrete/continuous variables; (2) the ability to effectively handle stochastic and deterministic constraints of various kinds; (2) the effective utilization of parallel computing at the linear algebra level, sample replication level, iteration level, as well as at the algorithmic level; (3) the effective handling of multiple simulation outputs; (4) the incorporation of performance measures other than expected values, such as risk;  (5) the continued consolidation of various techniques and their potential synergy in hybrid algorithms; (6) the use of automatic differentiation techniques in the estimation of simulation derivatives when possible; (7) the continued emphasis on providing guarantees of convergence to optima for local and global optimization routines in general settings; (8) the availability and ease of comparison of the performance of available approaches on different applications; and (9) the continued reflection of sophisticated methodology arising from the literature in commercial simulation packages.

\end{singlespacing}

\end{document}